\documentclass[lettersize,journal]{IEEEtran}
\usepackage{amsmath,amsfonts}
\usepackage{algorithmic}
\usepackage{algorithm}
\usepackage{array}
\usepackage[caption=false,font=normalsize,labelfont=sf,textfont=sf]{subfig}
\usepackage{textcomp}
\usepackage{stfloats}
\usepackage{url}
\usepackage{verbatim}
\usepackage{graphicx}
\usepackage{cite}

\usepackage{xcolor}
\usepackage{booktabs}
\usepackage{multirow}
\usepackage{makecell}

\begin{document}

\title{A Topology-Aware Positive Sample Set Construction and Feature Optimization Method in Implicit Collaborative Filtering}

\author{Jiayi Wu, Zhengyu Wu, Xunkai Li, Rong-Hua Li, and Guoren Wang
	\thanks{Jiayi Wu, Zhengyu Wu, Xunkai Li, Rong-Hua Li, and Guoren Wang are with
		the Beijing Institute of Technology, Beijing 100811, China (Correspondence to: Rong-Hua Li \textless lironghuabit@126.com\textgreater.)}
}


\markboth{Journal of \LaTeX\ Class Files,~Vol.~14, No.~8, August~2021}%
{Shell \MakeLowercase{\textit{et al.}}: A Sample Article Using IEEEtran.cls for IEEE Journals}


\maketitle

\begin{abstract}
Negative sampling strategies are widely used in implicit collaborative filtering to address issues like data sparsity and class imbalance. However, these methods often introduce false negatives, hindering the model’s ability to accurately learn users’ latent preferences. To mitigate this problem, existing methods adjust the negative sampling distribution based on statistical features from model training or the hardness of negative samples. Nevertheless, these methods face two key limitations: (1) over-reliance on the model’s current representation capabilities; (2) failure to leverage the potential of false negatives as latent positive samples to guide model in learning user preferences more accurately. 
To address the above issues, we propose a \underline{\textbf{T}}opology-aware \underline{\textbf{P}}ositive \underline{\textbf{S}}ample set \underline{\textbf{C}}onstruction and \underline{\textbf{F}}eature \underline{\textbf{O}}ptimization method (TPSC-FO). First, we design a simple topological community-aware false negative identification (FNI) method and observed that topological community structures in interaction networks can effectively identify false negatives. Motivated by this, we develop a topology-aware positive sample set construction module. This module employs a differential community detection strategy to capture topological community structures in implicit feedback, coupled with personalized noise filtration to reliably identify false negatives and convert them into positive samples. Additionally, we introduce a neighborhood-guided feature optimization module that refines positive sample features by incorporating neighborhood features in the embedding space, effectively mitigating noise in the positive samples. Extensive experiments on five real-world datasets and two synthetic datasets validate the effectiveness of TPSC-FO.
\end{abstract}


\section{Introduction}

\IEEEPARstart{I}{mplicit} feedback \cite{1} (e.g., clicks, browsing) has become the primary data source for modern recommender systems due to its accessibility and objectivity \cite{2}. Compared to explicit feedback \cite{3} (e.g., ratings, reviews), data generated from users' natural behaviors not only mitigates the sparsity issue of explicit feedback but is also easier to model \cite{4}. Collaborative filtering (CF) \cite{5}, as the most prevalent technique in recommender systems, effectively uncovers latent preference signals from implicit feedback through behavioral similarities \cite{6}. However, the large volume of unobserved data in implicit feedback poses challenges for model training, such as computational efficiency and gradient dilution \cite{7}. To address this, negative sampling strategies \cite{8} have emerged as a standard solution, which enhances both recommendation accuracy and computational efficiency in implicit CF by designing specific sample distributions to sample negatives from massive unobserved data.

\begin{figure}[t]
	\includegraphics[width=\linewidth]{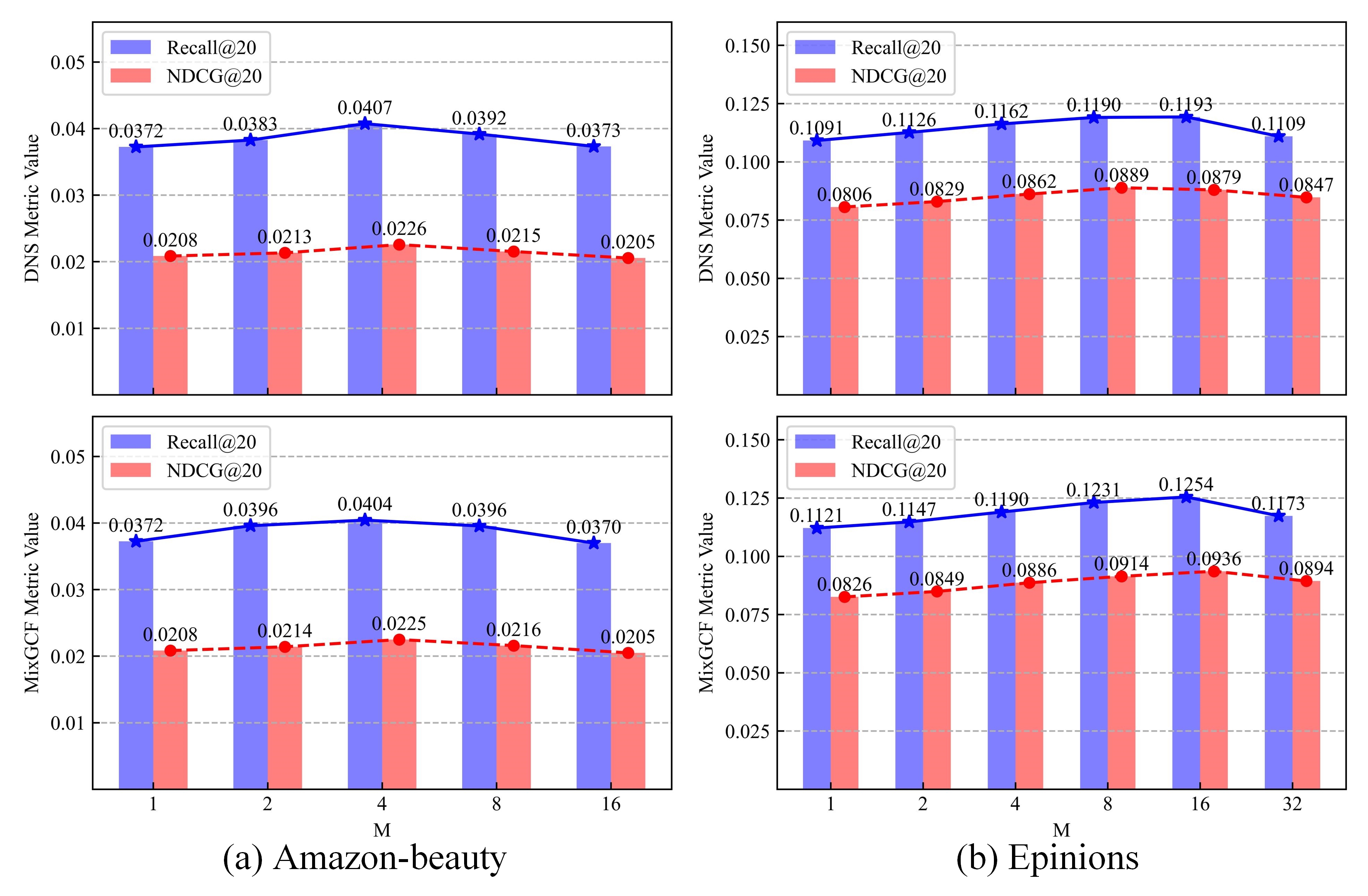}

	\caption{Impact of positive and negative sample similarity.}
	\label{fig: Impact of negative sample difficulty}
	
	\vspace{-0.2cm}
	
\end{figure}

Existing negative sampling methods\cite{9, 10, 11, 12} focus on constructing hard negative samples, prioritizing those with higher similarity to positive samples. Compared with random negatives, such samples yield larger gradient magnitudes that accelerate convergence and enhance the model’s discriminative ability. However, excessive similarity also increases the risk of false negatives, thereby degrading model performance \cite{13}. As shown in Fig.~1, we conducted experiments to verify this phenomenon. Specifically, we controlled the similarity between positive and negative samples by adjusting the candidate negative pool size $M$. A larger $M$ expands the range of candidate items, making it more likely to include samples that are closer to the positive ones. Similarity-based samplers such as DNS and MixGCF then select the hardest negatives, those with the highest similarity to the positives, from this enlarged pool. Consequently, increasing $M$ raises the average similarity of sampled negatives to the positives, which improves performance at first but eventually leads to more false negatives and a decline in recommendation accuracy. To mitigate this challenge, existing methods design negative sample distributions based on predictive variance \cite{13} or dynamically control negative sample difficulty \cite{14} to avoid samples that may be false negatives during negative sampling. However, these negative sampling methods generally have the following issues: (1) \textbf{Representation-dependent sampling}: They heavily rely on the quality of the model's current representations. When representation quality is low (e.g., due to insufficient training or data sparsity), sampling decisions based on these representations may increase the risk of false negatives. (2) \textbf{Missed false negative potential}: False negatives are essentially unexposed potential positive samples, and failing to utilize them causes the model to miss opportunities to accurately learn such preferences, ultimately affecting recommendation performance.

\begin{figure}[t]

	\includegraphics[width=\linewidth]{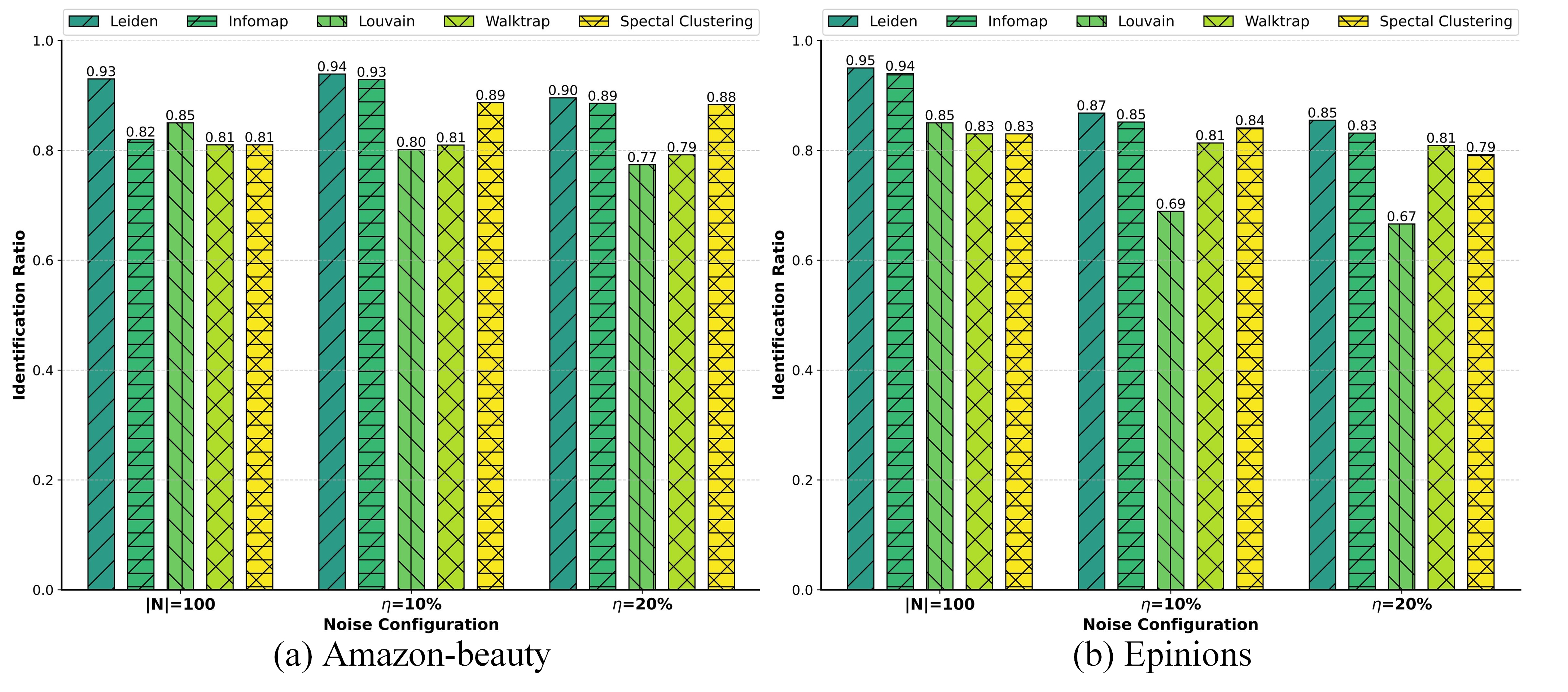}
	\caption{
		The performance of ComFNI on Amazon-beauty and Epinions.}
		
		\vspace{-0.2cm}
	
\end{figure}

To address the aforementioned issues, we innovatively propose a model-agnostic solution. We first consider that false negatives are essentially unexposed positive samples, potentially correlated with users' true preferences. Inspired by the "homophily principle" in complex network theory \cite{15, 16} (where nodes with similar features tend to form tightly connected communities), we design a topological community-aware false negative identification (FNI) method, named ComFNI. As illustrated in Algorithm 1, ComFNI first partitions the interaction network into communities using a community detection algorithm. Within each community, it computes the Cartesian product of users and items to construct candidate user–item pairs and identifies non-interacted pairs as the false negative set. This design allows ComFNI to flexibly integrate any community detection algorithm to capture the topological community structures of the interaction network for identifying false negatives. To verify the effectiveness of this method, we simulate false negatives by randomly removing 100, 10\%, and 20\% of positive samples from the training set, and then evaluate the ability of ComFNI using different community detection algorithms to identify these false negatives. As shown in Fig.~2, ComFNI achieves an FNI ratio exceeding 80\% in most cases, demonstrating that topological community features can effectively identify false negatives.

Building upon this, we propose a \underline{\textbf{T}}opology-aware \underline{\textbf{P}}ositive \underline{\textbf{S}}ample set \underline{\textbf{C}}onstruction and \underline{\textbf{F}}eature \underline{\textbf{O}}ptimization method (TPSC-FO), which includes a topology-aware positive sample set construction module and a neighborhood-guided feature optimization module. The topology-aware positive sample set construction module first constructs a candidate false negative set based on the consensus of FNI results obtained using different community detection algorithms in ComFNI. Then, it employs a personalized noise filtration to identify high-similarity samples within the candidate set as false negatives. Finally, these samples are incorporated into the positive sample set, transforming them from interfering items into positive samples with supervised signals. However, both newly introduced and original positive samples may contain noise. Existing denoising recommendation methods \cite{27,28,29,30} design their denoising strategies based on dataset-specific observations, and typically prune or reweight noisy samples. Such dataset-dependent designs limit their generalization ability, leading to positive supervisory signal sparsity and consequently degrading model performance. To address this issue, we design a neighborhood-guided feature optimization module that refines positive sample representations by integrating shared features of neighboring samples in the embedding space, effectively mitigating noise while preserving the strength of positive supervisory signals.

\begin{algorithm}[t]
	\renewcommand{\algorithmicrequire}{\textbf{Input:}}
	\renewcommand{\algorithmicensure}{\textbf{Output:}}
	\caption{Community-aware False Negative Identification method (ComFNI)}
	\label{alg: 1}
	\begin{algorithmic}[1]
		\REQUIRE The training set ${D_{train}} = \left\{ {U,I,R} \right\}$, Community detection algorithm $f\left(  \cdot  \right)$.
		
		\STATE $G = \left\langle {V,E} \right\rangle  \leftarrow $ Construct a user-item bipartite network based on  ${D_{train}}$
		\STATE $Se{t_f} \leftarrow \emptyset $
		\STATE ${C_f} = \left\{ {{c_1},{c_2},...,{c_m}} \right\} \leftarrow f\left( G \right)$
		\FOR{each \( c_j \in C_f \)}
		\STATE $\text{temp}_u \gets \{n{\in}c_j \mid n{\in}U\},\ \text{temp}_i \gets \{n{\in}c_j \mid n{\in}I\}$
		
		\STATE \( S \gets \{(u,i) \mid u\!\in\!temp_u,\quad i\!\in\!temp_i\} \) 
		\STATE \( S \gets \{ (u,i) \in S \mid {{r_{ui}} = 0} \} \) 
		\STATE \( Set_f \gets Set_f\!\cup S \) 
		
		\ENDFOR
		\ENSURE False negative set $Se{t_f}$
	\end{algorithmic} 
\end{algorithm}

This paper has made the following contributions:

\begin{itemize}

\item \textbf{New Perspective}. 
To the best of our knowledge, TPSC-FO is the first to introduce topological community features into negative sampling for implicit CF, with empirical analysis confirming their effectiveness in identifying false negatives.

\item \textbf{{New Method}}. 
TPSC-FO first constructs a false negative-empowered positive sample set through topological community features and personalized noise filtration, overcoming the limitations of traditional methods that heavily rely on model representation quality, and achieving effective transformation of noisy data into positive samples. Then, a neighborhood-guided feature optimization module is designed to effectively mitigate the impact of noise in positive samples.

\item \textbf{{SOTA Performance}}. 
Extensive experiments on 5 real-world and 2 synthetic datasets demonstrate that TPSC-FO achieves state-of-the-art performance and can be integrated into various implicit CF recommenders, working alongside different negative sampling strategies to further enhance model performance.

\end{itemize}

\begin{figure*}[tbp]
	\centering
	\includegraphics[width=\linewidth]{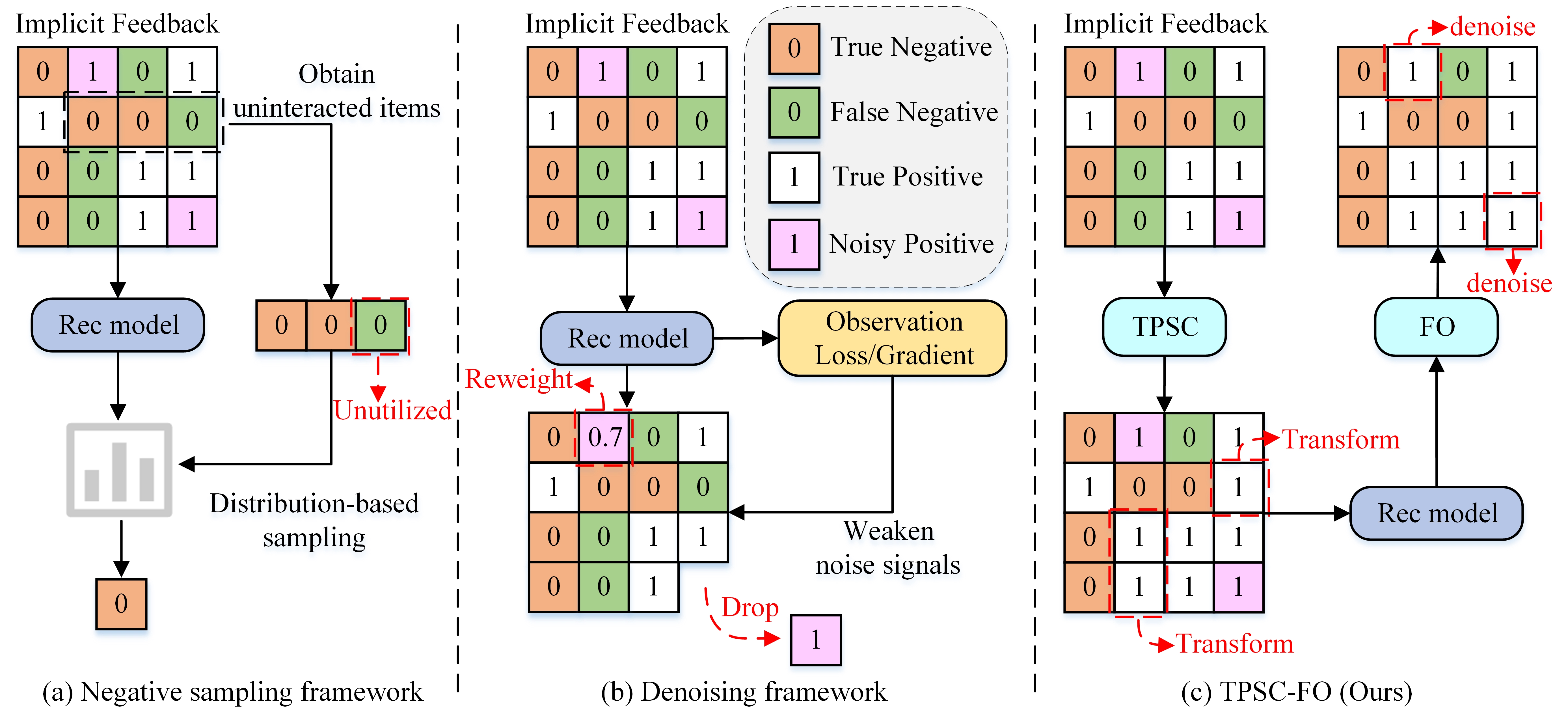}
	
	\vspace{-0.3cm}
	
	\caption{Illustration comparison: (a) Negative sampling heavily relies on the recommendation model’s representation quality. By selecting negative samples from uninteracted items, it fails to utilize the potential of false negatives. (b) Drop- and reweight-based denoising results in sparse positive supervisory signals. (c) TPSC-FO first employs the model-agnostic TPSC module to identify and transform false negatives into positive samples, then uses FO module to denoise positive samples, avoiding the issue of sparse positive supervisory signals.}

	\vspace{-0.2cm}
	
\end{figure*}

\section{Preliminaries}

\subsection{Background}

\textbf{Topological community structure} refers to a subgraph structure in complex networks composed of densely connected nodes, with significantly higher internal connection density than between subgraphs \cite{18, 19}. In recommender systems, these community structures is essentially a topological representation of user preference similarity, where users with similar interests spontaneously form dense connections with similar items \cite{20}. Therefore, accurately capturing these structures helps uncover latent user preferences, providing critical topological evidence for identifying false negatives and augmenting positive samples.

\vspace{0.15cm}

\hspace{-0.35cm}\textbf{Community detection algorithms} are widely used to capture topological community structures in networks and can be categorized based on their optimization approaches into iterative optimization methods based on deep learning and traditional heuristic methods. The former typically follow a two-stage framework of graph neural networks combined with clustering algorithms or end-to-end community affiliation learning to construct communities \cite{21}, but these methods have high computational complexity, with training times often exceeding those of downstream recommendation tasks. The latter, based on graph theory \cite{23} and information theory \cite{24}, directly define community partitioning criteria based on network topology without requiring parameter learning \cite{22}. Therefore, we selects five popular heuristic community detection algorithms (Leiden \cite{33}, Infomap \cite{34}, Louvain \cite{35}, Spectral Clustering \cite{36}, and Walktrap \cite{37}) to capture community structures. Below, we briefly introduce these algorithms. The Louvain \cite{35} algorithm employs a greedy strategy for hierarchical clustering, achieving modularity maximization through local node movements and community merging. The Leiden \cite{33} algorithm, as an improved version, enhances partition quality by enforcing community connectivity and optimizing merging strategies. Infomap \cite{34}, based on information theory, identifies communities by minimizing the encoding length of random walks. Spectral clustering \cite{36} utilizes the eigenvectors of the Laplacian matrix for dimensionality reduction followed by clustering. The WalkTrap \cite{37} algorithm generates communities through hierarchical clustering by analyzing the similarity of probability distributions from random walks.

\subsection{Related Work}

Most negative sampling methods focus on constructing high-quality hard negative samples \cite{10, 11, 12} to enhance model discriminability by increasing the similarity between positive and negative samples. However, these methods overlook the issue that overly similar negatives may actually be false negatives, impairing model's accuracy in learning user preferences. To address this, RecNS \cite{25} and RealHNS \cite{26} incorporate additional information (e.g., unclicked exposed data, cross-domain data) to reduce false negative risks, but its costly data collection limits generalizability. SRNS \cite{13} proposes a variance-based sampling strategy to avoid false negatives, which show higher prediction variance than hard negatives. AHNS \cite{14} controls negative sample difficulty based on positive sample prediction score to mitigate this issue. However, as illustrated in Fig.~3\:(a), these methods heavily rely on the recommendation model’s current representation quality, resulting in inconsistent false negative identification (FNI) for the same sample across different training stages. Additionally, they focus on avoiding false negatives during sampling, thereby failing to utilize their potential as positive supervisory signals. Motivated by these issues, as illustrated in Fig. ~3\:(c), we develop TPSC, a novel model-agnostic solution that introduces topological community features to identify false negatives and transforms them into positive samples, thereby constructing a false negative-empowered positive sample set. Extensive empirical study demonstrates the effectiveness of TPSC.

Although topological community features enable the construction of high-quality positive sample sets, both original and false negative-empowered sets may contain noise. To address this, T-CE \cite{27}, R-CE \cite{27}, DCF \cite{28}, and PLD \cite{29} denoise positive samples based on statistical features of loss or gradients. BOD \cite{30} and SGDL \cite{31} employ a bi-level optimization architecture to learn denoising weight matrices. DeCA \cite{32} performs denoising based on prediction consistency across different recommendation models. However, as illustrated in Fig. ~3\:(b), these methods prune or reweight noisy samples based on observed dataset-specific statistical patterns or optimization heuristics, which limits their generalizability and degrades model performance (see Table II). More critically, low-accuracy denoising results in positive supervisory signal sparsity, thereby limiting the model’s ability to learn user preferences, since positive samples constitute the sole supervisory signal guiding preference learning. Driven by this issue, we develop a neighborhood-guided feature optimization module, as illustrated in Fig.~3\:(c), which effectively mitigates positive sample noise without weakening the positive supervisory signal. Extensive experiments verify its effectiveness.

\begin{figure}[t]
	\includegraphics[width=\linewidth]{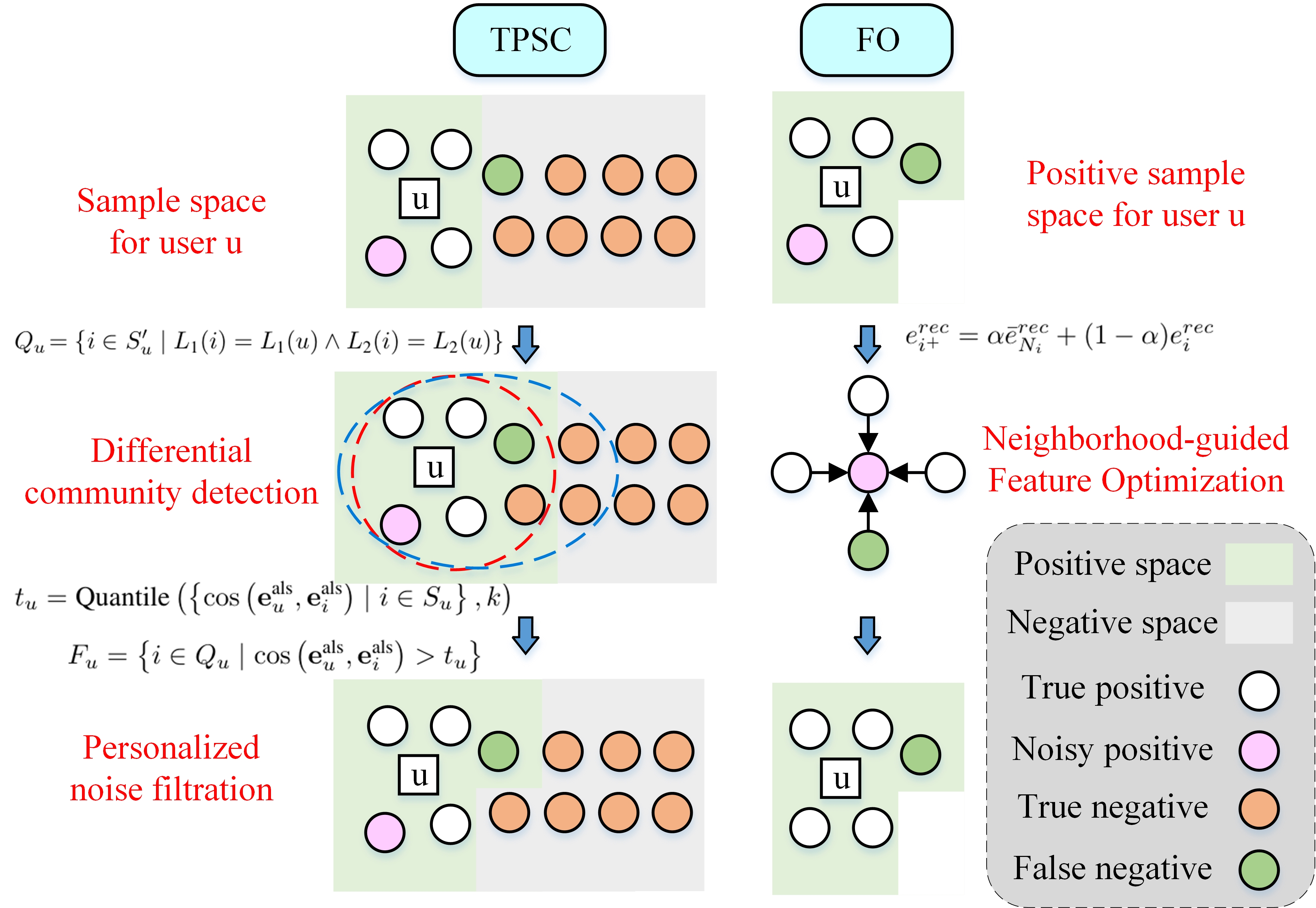}
	\caption{Impact of positive and negative sample similarity.}
	\label{fig: Impact of negative sample difficulty}
	
\end{figure}

\section{The Proposed Method}

In this section, we first formalize the problem. Then, we introduce our proposed method TPSC-FO, whose workflow is detailed in Fig.~4. It comprises a topology-aware positive sample set construction module (TPSC) and a neighborhood-guided feature optimization module (FO). Finally, we analyze its time complexity and validate the feature optimization’s effectiveness via theoretical analysis.

\subsection{Problem Formalization}

Given an implicit feedback dataset ${D_{train}} = \left\{ {U,I,R} \right\}$, where $U$ and $I$ represent the user and item set, respectively, and $R$ is a binary matrix $R \in {\left\{ {0,1} \right\}^{|U| \times |I|}}$, with ${r_{ui}} = 1$ indicating an interaction (e.g., click, view, purchase) between user $u$ and item $i$, and $ r_{ui} = 0 $ indicating no interaction. For any user $u \in U$, the positive sample set is defined as ${S_u} = \left\{ {i \in I|{r_{ui}} = 1} \right\}$, and the unobserved item set as ${S'_u} = \left\{ {i \in I|{r_{ui}} = 0} \right\}$. The goal of implicit CF is to train a recommendation model to predict the preference probability ${\hat r_{ui}} = f\left( {u,i|\Theta } \right)$ between user $u$ and an uninteracted item $i$. We train the model by minimizing the Bayesian Personalized Ranking (BPR) \cite{38} loss:
\begin{equation}
	\label{eq:bpr_loss}  
	\ell_{\text{BPR}}(u,S_u,S'_u) = 
	\!	-\!\!\! \sum_{\substack{
			u \in {U}, 
			i \in S_u,\\ 
			j \in \text{NS}(S'_u)
	}} 
	\ln \sigma(\hat{r}_{ui} - \hat{r}_{uj}) 
	+ \lambda \|\Theta\|_2^2,
\end{equation}
where ${\hat r_{ui}} = e_u^T{e_i}$ denotes the preference probability of $u$ for $i$, while ${e_u}$ and ${e_i}$ represent the embeddings of $u$ and $i$, respectively. $j \in \text{NS}\left( {{S'_u}} \right)$ denotes the negative sample $j$ sampled from ${S'_u}$ via the negative sampling method $\text{NS}\left(  \cdot  \right)$. In real-world scenarios, ${S'_u}$ consists of true negative samples ${T_u}$ and false negatives ${F_u}$. True negative samples represent items in  ${S'_u}$ that $u$ is not interested in, while false negatives represent samples in ${S'_u}$ that reflect $u$'s preferences but are unexposed. Training with such ${S'_u}$ leads to bias in model preference estimation. Therefore, we need to identify ${F_u}$ within it and convert them into positive supervision signals to guide model training. The task of training a recommender using a false negative-empowered positive sample set is defined as follows:
\begin{equation}
	\Theta^* = \min \mathcal{L}_{\text{BPR}}\left(u, S_u^+, {S'_u}^-\right),
\end{equation}
where ${S'_u}^ -  = {S'_u}\backslash {F_u}$, $S_u^+  = {S_u} \cup {F_u}$. ${S'_u}^ -$ and $S_u^+$ represent $u$'s true negative sample set and false negative-empowered positive sample set, respectively.

\subsection{Topology-aware Positive Sample Set Construction}

To accurately identify ${F_u}$, inspired by Fig.~2, we propose a differential community detection strategy. This strategy first models implicit feedback as a user-item bipartite network and then constructs a candidate false negative set based on the consensus results of ComFNI using different community detection algorithms. Specifically, considering the accuracy of false negative identification (See Fig.~2) and time complexity, we adopt Leiden based on modularity maximization and Infomap based on information flow compression for community partitioning of the bipartite network. For any user $u$ and uninteracted item $i$, $i$ is included in the candidate false negative set ${Q_u}$ if it satisfies: 1) $i$ belongs to the same community as $u$ in the Leiden partitioning; 2) $i$ shares the same community label as $u$ in the Infomap partitioning. The formulation is as follows:
\begin{equation}
	Q_u\! = \left\{ i \in S'_u \mid L_{\text{ld}}(i) = L_{\text{ld}}(u) \land L_{\text{im}}(i) = L_{\text{im}}(u) \right\},
\end{equation}
where $L_{\text{ld}}(i)$ and $L_{\text{im}}(i)$ denote the community labels of $i$ in the Leiden and Infomap partitionings, respectively. This strategy enhances topological consistency and reduces noise from individual algorithms by leveraging the consensus of community detection algorithms from different perspectives (information flow and modularity), thus ensuring strong correlation between candidate false negatives and users in the global topological structure.

However, most community detection algorithms, such as Louvain and Leiden (which maximize modularity) and Infomap (which minimizes description length), suffer from the issue of forcibly merging small-scale interest communities \cite{39}. This results in noise samples in ${Q_u}$ that are topologically similar but differ in interest preferences. To address this, we propose personalized noise filtration. First, to achieve fast node embedding learning, we adopt the Alternating Least Squares \cite{40} (ALS) matrix factorization model to learn user and item embeddings $ e_{\text{als}}^U $, $ e_{\text{als}}^I $. Compared to other matrix factorization models, such as SVD \cite{41} and NMF \cite{42}, ALS exhibits superior performance and efficiency (see Table IV). Then, we compute the cosine similarity between target user $ u $ and all interacted items, setting a personalized threshold $ t_u $ to control the strictness of false negative filtration, calculated as follows:
\begin{equation}\label{eq:threshold}
	t_u = \text{Quantile}\left( \left\{ \cos\left( \mathbf{e}_u^\text{als}, \mathbf{e}_i^\text{als} \right) \mid i \in S_u \right\}, k \right),
\end{equation}
where $\left\{ {cos\left( {e_u^{als},e_i^{als}} \right)|i \in {S_u}} \right\}$ denotes the set of similarities between $u$ and interacted items, and $Quantile\left( {set,k} \right)$ denotes the value at the $k$-th percentile of $set$. For any item $i \in {Q_u}$, if the similarity between $u$ and $i$ exceeds $t_u$, then $i$ is deemed a valid false negative.
\begin{equation}
	F_u = \left\{ i \in Q_u \mid \cos\left( \mathbf{e}_u^\text{als}, \mathbf{e}_i^\text{als} \right) > t_u \right\}.
\end{equation}
Finally, we combine the filtered false negative set $F_u$ with the original positive sample set $S_u$ to obtain a topology-aware positive sample set $S_u^+$. This process transforms noisy negative samples into positive supervision signals, providing richer training data for subsequent user preference learning, effectively mitigating the data sparsity issue.

\begin{algorithm}[t]
	\renewcommand{\algorithmicrequire}{\textbf{Input:}}
	\renewcommand{\algorithmicensure}{\textbf{Output:}}
	\caption{Topology-aware Positive Sample Set Construction}
	\label{alg: 2}
	\begin{algorithmic}[1]
		\REQUIRE The training set ${D_{train}}$, Positive sample set $S_U$, $k$.
		
		\STATE $se{t_{ld}} \leftarrow \text{ComFNI}\left( {{D_{train}},leiden\left(  \cdot  \right)} \right)$;
		\STATE $se{t_{im}} \leftarrow \text{ComFNI}\left( {{D_{train}},infomap\left(  \cdot  \right)} \right)$;
		\STATE Construct candidate false negative set $Q_u$ based on $se{t_{ld}}$ and $se{t_{im}}$ by Eq.(3);
		\STATE Obtain user and item embeddings $e_U^{als},e_I^{als}$ by ALS;
		\FOR {each \( u \in U \)}
		\STATE ${F_u} \leftarrow \emptyset $;
		\STATE $temp \gets \{ \cos(e_u^{als}, e_i^{als}) \mid i \in S_u \}$;
		\STATE Compute threshold $t_u$ based on $temp$ by Eq.(4).
		\FOR {each \( (u,i) \in Q_u \) }
		\IF{$\cos \left( {e_u^{als},e_i^{als}} \right) > t_u$}
		\STATE ${F_u} \leftarrow {F_u} \cup i$;
		\ENDIF
		\ENDFOR
		\ENDFOR		
		\STATE ${S_U}^ +  = {S_U} \cup {F_U}$, ${S'_U}^ -  = {S'_U}\backslash {F_U}$
		\ENSURE Topology-aware positive sample set $S_U^+$.
	\end{algorithmic} 
\end{algorithm}

Algorithm 2 describes the construction of a topology-aware positive sample set. (Lines 1-3) A candidate false negative set is built based on topological community features. (Lines 4-14) Personalized noise filtration is applied to screen the candidate false negative set, yielding the final false negative set. (Line 15) The topology-aware positive sample set is constructed by merging the final false negative set and original positive sample set.

\begin{algorithm}[t]
	\renewcommand{\algorithmicrequire}{\textbf{Input:}}
	\caption{The training process with TPSC-FO}
	\label{alg:3}
	\begin{algorithmic}[1]
		\REQUIRE The training set $D_{train}$, Implicit CF model $\text{rec}\left(  \cdot  \right)$, Negative sampling method $\text{NS}\left(  \cdot  \right)$.
		\STATE Obtain the positive sample set ${S_U}$ and the unobserved item set ${S'_U}$ for all users based on $D_{train}$.
		\STATE Call algorithm \ref{alg: 2} to obtain topology-aware set $S_U^+$.
		\STATE ${D_{pairs}} \leftarrow \bigcup\nolimits_{u \in U} {\left\{ {\left( {u,i} \right)\left| {i \in S_u^ + } \right.} \right\}}$.
		\FOR{$t = 1, 2,…,T$}
		\STATE Sample a mini-batch of pairs $D_{batch}$ from $D_{pairs}$.
		\STATE $\ell  = 0$.
		\FOR{each $\left( {u,i} \right) \in D_{batch}$}
		\STATE Randomly select $n$ positive samples from $S_u^ + \backslash \left\{ i \right\}$ as the neighborhood sample set ${N_i}$ for $(u, i)$.
		\STATE $e_u^{rec},e_i^{rec},e_{{N_i}}^{rec} \leftarrow $ Obtain the embeddings of $u$, $i$, $N_i$ by $\text{rec}\left(  \cdot  \right)$.
		\STATE Obtain the denoised positive sample embedding $e_{{i^ + }}^{rec}$ based on $e_{{N_i}}^{rec}$ and $e_i^{rec}$ by Eq.(6).
		\STATE Obtain negative item embeddings ${e_{i^ - }^{rec}}$ by $\text{NS}\left(  \cdot  \right)$.
		\STATE $\ell  = \ell  + \log \sigma \left( {e_{{i^ - }}^{rec} \cdot e_u^{rec} - e_{{i^ + }}^{rec} \cdot e_u^{rec}} \right)$.
		\ENDFOR
		\STATE Update the parameters of the implicit CF model $\text{rec}\left(  \cdot  \right)$ by descending the gradients ${\nabla _\theta }\ell $.
		\ENDFOR
	\end{algorithmic}
	
\end{algorithm}

\subsection{Neighborhood-guided Feature Optimization}

The topology-aware positive sample set $S_u^+$, built upon the original positive sample set $S_u$, incorporates a latent positive sample set $F_u$ constructed based on topological consistency and feature similarity. However, this expansion increases the risk of $S_u^+$ being affected by noise, as noise may originate not only from $S_u$ (e.g., unconscious clicks or accidental interactions by users) but also from $F_u$ (due to biases introduced by partition granularity and similarity threshold).

To mitigate the impact of these noises, we design a neighborhood-guided positive sample feature optimization module. First, for any user $u$ and item $i \! \in \! S_u^ + $, we randomly select $ n $ samples from $S_u^ + \backslash \left\{ i \right\}$ as the neighborhood positive sample set ${N_i} = \left\{ {{j_1},{j_{2,...,}}{j_n}} \right\}$ for the interaction data $ (u, i) $. Next, through forward propagation of the recommendation model, we obtain the neighborhood positive sample embedding set $e_{{N_i}}^{{\mathop{\rm re}\nolimits} c} = \left\{ {e_{{j_1}}^{{\mathop{\rm re}\nolimits} c},e_{{j_2}}^{{\mathop{\rm re}\nolimits} c},...,e_{{j_n}}^{{\mathop{\rm re}\nolimits} c}} \right\}$. Considering computational efficiency and potential noise in $N_i$, we aggregate the neighborhood positive sample embedding set using mean pooling to smooth out noise from individual samples and extract common user preference embeddings $\bar e_{{N_i}}^{rec}$. Finally, we employ the mixup \cite{43} technique to integrate the common embeddings $\bar e_{{N_i}}^{rec}$ into the original item embeddings $e_i^{{\mathop{\rm re}\nolimits} c}$, yielding denoised embeddings, denoted as:
\begin{equation}
	e_{i^+}^{rec} = \alpha \bar{e}_{N_i}^{rec} + (1 - \alpha) e_i^{rec},
\end{equation}
where $\alpha \! \sim \! \text{U}\left( {0,1} \right)$ controls a diversified fusion ratio to enhance the robustness of embeddings. Notably, unlike previous methods \cite{10} that use mixup between positive and negative sample features to construct hard negative sample features, our approach innovatively applies mixup to combine positive sample feature with its neighborhood positive sample features, focusing on denoising positive sample features to enable the model to learn user preferences more accurately.

To illustrate how the neighborhood-guided feature optimization (FO) module mitigates positive sample noise, we analyze its effect on embedding expectation and variance.
Consider a positive pair $(u, i)$, where $e_u$ and $e_i$ denote the embeddings of the user and item, and $e_{N_i}$ represents the embeddings of $i$’s positive-sample neighbors of size $n$. 
Each positive embedding $ e_i $ comprises a embedding $e_i^{true}$ reflecting user's true preferences and a noise embedding ${\varepsilon_i}$, denoted as ${e_i} = e_i^{true} + {\varepsilon_i}$. This noise primarily arises from random perturbations (e.g., accidental clicks, data sparsity), which converge to a Gaussian distribution under the Central Limit Theorem \cite{44}. Thus, we let ${\varepsilon_i} \sim N\left( {0,\sigma^2} \right)$. The optimized embedding of item $i$ is expressed as follows:
\begin{equation}
	\begin{aligned}
		e_{i^+} &= \alpha \bar{e}_{N_i} + (1 - \alpha) e_i \\
		&= \frac{\alpha}{n}\sum_{j \in N_i} \left( e_j^{true} + \varepsilon_j \right) + (1 - \alpha)\left( e_i^{true} + \varepsilon_i \right).
	\end{aligned}
\end{equation}
Since both $\bar e_j^{true}$ and $e_i^{true}$ reflect the true preferences of $u$, we can obtain $E\left( {\bar e_j^{true}} \right) \! \approx \! E\left( {e_i^{true}} \right)$. Based on this, the optimized feature expectation is given by:
\begin{equation}
	E[e_{i^+}] \approx E[e_i^{true}],
\end{equation}
which means that the optimized positive sample embeddings still reflect users' true preferences.

Given that ${\varepsilon _i}$ is the primary factor causing positive sample embedding to deviate from user's true preferences, the goal of denoising is to reduce the impact of ${\varepsilon _i}$. In contrast, the variance of $e_i^{true}$ reflects the preference differences among users, so we focus on computing the variance of the optimized noise embedding $\varepsilon _i^ + $:
\begin{equation}
	Var[\varepsilon_i^+] = \left(\frac{\alpha^2}{n} + (1 - \alpha)^2\right)\sigma_i^2 < \sigma_i^2.
\end{equation}
Since $\alpha \sim \text{U}(0,1)$, it follows that $0 < \alpha < 1$; and because $n$ represents the size of the neighborhood positive sample set, $n$ is a constant greater than 1. Therefore, this inequality always holds, indicating that the neighborhood-guided feature optimization module effectively reduces the variance of noise in the positive sample embeddings.

The above results demonstrate that FO module effectively reduces the variance of the noise embeddings while preserving the expectation of the embeddings reflecting users' true preferences, achieving denoising optimization of the positive sample embeddings.

Algorithm 3 describes the overall workflow of TPSC-FO. (Lines 1–2) construct the topology-aware positive sample set. (Lines 3–15) train the implicit CF model using TPSC-FO, where (lines 8–10) apply the neighborhood-guided feature optimization module to denoise the positive sample.

\subsection{Time Complexity}

The time complexity of TPSC-FO primarily stems from two parts: the topology-aware positive sample set construction module in the data preparation phase and the neighborhood-guided feature optimization module in the model training phase. The time complexity of the former involves community detection, node embedding learning, and noise filtration. The time complexity of community detection using Leiden and Infomap in sparse graph scenarios can be approximated as $O\left( {V\log V} \right)$, where $V$ represents the number of nodes. The time complexity of obtaining node embeddings through ALS is $O\left( {TE{d^2}} \right)$, where $ T $ denotes the number of iterations, $ E $ represents the number of interactions, and $ d $ indicates the embedding dimension. The time complexity of personalized noise filtration primarily arises from computing the cosine similarity of all interactions, expressed as $O\left( {Ed} \right)$. Therefore, the complexity of constructing the topology-aware positive sample set is $O\left( {TE{d^2} + V\log V} \right)$. The time complexity of the latter primarily involves neighborhood positive sample selection, mean pooling, and mixup operations. The complexity of neighborhood positive sample selection is $O\left( {{E^ + }n} \right)$, where $ n $ denotes the size of the neighborhood positive sample set and $E^ + $ represents the number of interactions in the topology-aware positive sample set. Mean pooling requires operating on the neighborhood positive sample set for each interaction, with a complexity of $O\left( {{E^ + }nd} \right)$. Mixup is applied to denoise each positive sample, with a complexity of $O\left( {{E^ + }d} \right)$. Thus, the time complexity of the positive sample feature optimization method is $O\left( {{E^ + }nd} \right)$. The overall complexity of TPNS is $O\left( {TE{d^2} + V\log V + {E^ + }nd} \right)$.

\section{Experiments}

In this section, we perform extensive experiments to evaluate TPSC-FO and answer the following research questions: \textbf{(RQ1)} How does TPSC-FO perform compared to previous negative sampling methods and denoising methods? How does each module in TPSC-FO contribute to the overall performance? \textbf{(RQ2)} How do the hyperparameters of TPSC-FO affect model performance? \textbf{(RQ3)} How do the choices of community detection algorithms and node embedding methods within the TPSC module affect the model’s effectiveness? \textbf{(RQ4)} How does TPSC-FO perform when integrated into other implicit CF models (e.g., matrix factorization \cite{45,53}.) and negative sampling methods?

\begin{table}[tbp]
	\caption{Dataset statistics.}
	\centering
	\resizebox{\linewidth}{16mm}{
		\setlength{\tabcolsep}{1mm}{
			\begin{tabular}{cccccc}
				\toprule
				Category & Dataset & \# Users & \# Items & \# Interactions & Density \\
				\midrule
				\multirow{4}{*}{\makecell{Real-world \\ dataset}} 
				& Amazon-beauty & 7.7k & 10.9k & 82.3k & 0.099\% \\
				& Amazon-home & 23.6k & 22.4k & 230.4k & 0.044\% \\
				& Epinions & 11.5k & 11.7k & 327.9k & 0.245\% \\
				& Gowalla & 29.9k & 40.1k & 1027.3k & 0.084\% \\
		        & Tmall & 47.9k & 41.4k & 2619.4k & 0.132\% \\
				\midrule
				\multirow{2}{*}{\makecell{Synthetic \\ noisy dataset}} 
				& Beauty-20\% & 7.7k & 10.9k & 75.9k & 0.091\% \\
				& Epinions-20\% & 11.5k & 11.7k & 276.4k & 0.206\% \\
				\bottomrule
			\end{tabular}
		}
	}
\end{table}

\subsection{Experimental Setup}

\textbf{Datasets and metrics.}
We evaluated TPSC-FO using five real-world datasets (Amazon-beauty \cite{46}, Amazon-home \cite{46}, Epinions \cite{48}, Gowalla \cite{11}, and Tmall \cite{52}) and two synthetic noisy datasets (Amazon-beauty-\%k and Epinions-\%k). Table I summarizes the statistics of these datasets. For the noisy datasets, we randomly removed k\% of the training data to simulate false negatives. For the real-world datasets, Amazon-home and Amazon-beauty are derived from user review data in the Home\&Kitchen and Baby product categories on the Amazon e-commerce platform. The Epinions dataset consists of user rating data from the Epinions website. The Gowalla dataset comprises user check-in data from the Gowalla social platform. The Tmall dataset contains user–item interaction records collected from the Tmall online retail platform. All datasets followed the same splitting methodology as previous studies\cite{46, 48, 11, 52}. For Amazon-home and Amazon-beauty, data before April 1, 2014, was used as the training set, with the remaining data as the test set, and 10\% of the training set interactions were randomly sampled as the validation set. The Epinions dataset was split into training, test, and validation sets in a 7:1:2 ratio. The Gowalla and Tmall datasets were divided into training, test, and validation sets in an 8:1:1 ratio. We adopt the widely used Recall@k and NDCG@k metrics in the recommendation field to evaluate the performance of recommender, where the value of $k$ = [10, 20], denoted as R@10, R@20, and N@20. The higher the value of the above metrics, the better the performance of the recommender system.

\textbf{Baselines.}
TPSC-FO effectively reduces the risk of false negatives in negative sampling through the TPSC module. To demonstrate its superiority, we compared it with six representative negative sampling methods: the commonly used RNS \cite{38}, three hard negative sampling methods (DNS \cite{9}, MixGCF \cite{10}, and DNS(M,N) \cite{12}), and two methods addressing false negatives (SRNS \cite{13} and AHNS \cite{14}). Additionally, since TPSC-FO denoises positive sample vectors through the FO module, we also compared it with five competitive denoising recommendation methods: T-CE \cite{27}, R-CE \cite{27}, DeCA \cite{32}, DCF \cite{28}, and PLD \cite{29}. To ensure a fair comparison, for all baselines focusing on negative samples, we use the user’s 1-hop item neighbors to obtain positive samples. For all methods focusing on positive samples (including ours), we employ a random negative sampling strategy \cite{38} to obtain negative samples. The salient characteristics of all baselines are outlined below:

\begin{itemize}
	\item RNS \cite{38}: A negative sampling method that randomly samples from uninteracted negative samples.
	\item DNS \cite{9}: A negative sampling method that dynamically selects hard negative samples based on the model's current prediction results.
	\item DNS(M,N) \cite{12}: An extended version of DNS that controls the difficulty of sampled negative samples through predefined parameters.
	The parameters $M$ and $N$ are set to 100 and 5 for all datasets, respectively.
	\item MixGCF \cite{10}: A negative sampling method based on mixup \cite{43} to construct hard negative samples.
	\item SRNS \cite{13}: A prediction variance-based method for filtering false negatives in negative sampling. The parameters $S_1$ and $varset\_size$ are set to 20 and 3,000, respectively.
	\item AHNS \cite{14}: The state-of-the-art negative sampling method that adaptively adjusts the hardness of negative samples based on positive samples, thereby mitigating the risk of false negatives. The parameters $\alpha$ and $\beta$ are set to 1.0 and 0.1 across all datasets. The parameter $p$ is set to 2, –2, 2, 2, and 2 for the Amazon-Beauty, Amazon-Home, Epinions, Gowalla, and Tmall datasets, respectively.
	\item T-CE \cite{27}: A denoising recommendation method that drops noisy samples based on sample losses. The parameters $\epsilon_N$ and $\epsilon_{max}$ controlling the drop rate are set to 30,000 and 0.1 for all datasets, respectively.
	\item R-CE \cite{27}: A denoising recommendation method that reweights positive samples based on sample losses. The parameter $\beta$ is set to 0.1 for all datasets.
	\item DeCA \cite{32}: A denoising recommendation method leveraging the prediction consistency of different recommendation models. This method employs the MF pre-trained model. The parameters $C_1$, $C_2$, and $\alpha$ are set to 10, 1000, and 1.0 across all seven datasets.
	\item DCF \cite{28}: A denoising recommendation method that denoises data through Sample Dropping Correction and Progressive Label Correction. The parameters $time\_interval$ and $relabel$ $ratio$ are set to 3 and 0.05 for all datasets, respectively.
	\item PLD \cite{29}: The state-of-the-art denoising recommendation method. Resampling positive samples based on user’s personal loss distribution. The parameters $\tau$ and $k$ are set to 0.1 and 5, respectively, across all seven datasets.
	
\end{itemize}

\textbf{Parameter settings.}
We conducted all experiments on LightGCN \cite{49}. Initial embeddings were initialized with the Xavier method \cite{50} at a dimension of 64. We employed the Adam optimizer \cite{51} with a learning rate of 0.001, set the mini-batch size to 2048, used an L2 regularization parameter of 0.0001, and the model layers is set to 3.
The parameters for the community detection algorithms were selected based on modularity. The $resolution\_parameter$ for Leiden was uniformly set to 0.01, and the hierarchical structure for Infomap was consistently set to two-level across all datasets.

\textbf{Experiment environment.}
Our experiments are conducted on a Linux server with Intel(R) Xeon(R) Gold 6240 CPU@2.60GHz, 251G RAM and 4 NVIDIA A100 80GB PCIe. Our proposed TPSC-FO is implemented in Python 3.10.13, torch 2.4.0, scikit-learn 1.6.1, leidenalg 0.10.2, infomap 2.8.0, implicit 0.5.2, and CUDA 12.4.0.

\textbf{Remark.}
To ensure a fair comparison, unless otherwise specified, the pool size for all methods that use candidate sample pool techniques is set to 10. All experiments take data leakage into account, and data contained in the validation and test set are removed from the topology-aware positive sample set. To ensure the reproducibility of the experiment, we set the seed to 2022. Our source code is available in the anonymous repository https://anonymous.4open.science/r/TPSC-FO-7D14/.

\begin{table*}[t]
	\centering
	\setlength{\abovecaptionskip}{0.3cm}
	\caption{Performance comparison on five real-world datasets (\%). The best result is \textbf{bold}.}
	\resizebox{\linewidth}{30mm} {
		\setlength{\tabcolsep}{2.0mm}{
			\begin{tabular}{c|ccc|ccc|ccc|ccc|ccc}
				\toprule
				\multirow{1}{*}{Dataset} & \multicolumn{3}{c|}{Amazon-beauty} & \multicolumn{3}{c|}{Amazon-home} & \multicolumn{3}{c|}{Epinions} & \multicolumn{3}{c|}{Gowalla} & \multicolumn{3}{c}{Tmall} \\ 
				\cmidrule(r){1-1} \cmidrule(r){2-4} \cmidrule(r){5-7} \cmidrule(r){8-10} \cmidrule(r){11-13} \cmidrule(r){14-16}
				\multirow{1}{*}{Method} & R@10 & R@20 & N@20 & R@10 & R@20 & N@20 & R@10 & R@20 & N@20 & R@10 & R@20 & N@20 & R@10 & R@20 & N@20\\
				\midrule
				RNS        & 2.43 & 3.84 & 2.11 & 1.08 & 1.78 & 0.88 & 7.09 & 11.03 & 8.07 & 13.59 & 19.76 & 12.35 & 4.45& 7.10& 4.96\\
				DNS        & 2.56 & 3.82 & 2.15 & 1.04 & 1.58 & 0.87 & 7.72 & 12.00 & 8.84 & 13.64 & 19.64 & 12.45& 4.66& 7.53& 5.26\\
				MixGCF     & 2.60 & 4.00 & 2.19 & 1.10 & 1.74 & 0.92 & 8.17 & 12.48 & 9.30 & 13.93 & 19.97 & 12.58&4.95& 8.00& 5.61\\
				DNS(M,N)   & 2.59 & 4.04 & 2.14 & 1.05 & 1.61 & 0.88 & 7.75 & 11.58 & 8.77 & 13.89 & 19.89 & 12.56&4.97& 7.90& 5.59\\
				SRNS       & 2.59 & 3.99 & 2.16 & 1.14 & 1.73 & 0.90 & 6.59 & 10.51 & 7.52 & 12.67 & 18.52 & 11.47&4.57& 7.34&5.16\\
				AHNS       & 2.59 & 3.94 & 2.19 & 1.21 & 2.07 & 1.00 & 7.51 & 11.66 & 8.62 & 13.66 & 19.65 & 12.37&4.94& 8.01& 5.59\\
				\cmidrule(r){1-16}
				T-CE       & 2.50 & 3.76 & 2.09 & 1.10 & 1.66 & 0.86 & 6.84 & 10.64 & 7.58 & 9.67 & 14.29 & 8.84&2.40& 4.14& 2.73\\
				R-CE       & 2.51 & 3.81 & 2.07 & 1.01 & 1.59 & 0.83 & 6.61 & 10.56 & 7.43 & 8.42 & 12.60 & 7.58&3.15& 5.28& 3.54\\
				DeCA       & 2.46 & 3.75 & 2.15 & 1.03 & 1.68 & 0.86 & 7.27 & 11.33 & 8.20 & 9.73 & 14.22 & 8.89&3.49& 5.86& 3.95\\
				DCF        & 2.50 & 3.78 & 2.10 & 1.04 & 1.62 & 0.84 & 7.12 & 11.15 & 8.06 & 9.63 & 14.29 & 8.80&3.07& 5.13& 3.46\\
				PLD        & 2.27 & 3.41 & 1.89 & 1.05 & 1.69 & 0.85 & 7.72 & 11.68 & 8.55 & 9.84 & 14.50 & 8.86&3.98& 6.48& 4.46\\
				\cmidrule(r){1-16}
				FO         & 2.47 & 3.93 & 2.19 & 1.21 & 2.07 & 1.01 & 8.03 & 12.11 & 8.96 & 13.89 & 20.07 & 12.62&5.32& 8.43& 5.98\\
				TPSC       & 2.75 & 4.03 & 2.31 & 1.22 & 1.87 & 1.02 & 9.13 & 12.90 & 11.22 & 17.40 & 23.24 & 16.15&8.04& 10.57& 9.75\\
				TPSC-FO    & \textbf{2.83} & \textbf{4.26} & \textbf{2.40} & \textbf{1.35} & \textbf{2.09} & \textbf{1.10} & \textbf{9.80} & \textbf{13.60} & \textbf{11.83} & \textbf{17.62} & \textbf{23.48} & \textbf{16.39}&\textbf{8.43}&\textbf{10.98}&\textbf{10.38}\\
				\bottomrule
			\end{tabular}
	}}
\end{table*}

\begin{table}[t]
	\centering
	\setlength{\abovecaptionskip}{0.2cm}
	\setlength{\belowcaptionskip}{-0.2cm}
	\caption{Performance comparison on synthetic datasets (\%).}
	\resizebox{\linewidth}{28mm} {
		\setlength{\tabcolsep}{1.2mm}{
			\begin{tabular}{c|cccc|cccc}
				\toprule
				\multirow{1}{*}{Dataset} & \multicolumn{4}{c|}{Amazon-beauty-20\%} & \multicolumn{4}{c}{Epinions-20\%} \\ 
				\cmidrule(r){1-1}\cmidrule(r){2-5} \cmidrule(r){6-9}
				\multirow{1}{*}{Method}& R@10 & N@10 & R@20 & N@20 & R@10 & N@10 & R@20 & N@20 \\
				\midrule
				RNS        & 2.23& 1.49& 3.32& 1.85&                    6.36& 5.89& 10.12& 7.24\\
				DNS        & 2.12& 1.45& 3.43& 1.87&                    6.97& 6.48& 10.90& 7.88\\
				MixGCF     & 2.26& 1.52& 3.57& 1.94&                    7.39& 6.87& 11.30& 8.24\\
				DNS(M,N)   & 2.19& 1.44& 3.58& 1.89&                    6.63& 6.20&  9.97& 7.39\\
				SRNS       & 2.23& 1.53& 3.63& 1.98&                    6.43& 5.67& 10.07& 7.01\\
				AHNS       & 2.33& 1.57& 3.49& 1.95&                    6.93& 6.45& 10.75& 7.82\\
				\cmidrule(r){1-9} 
				T-CE       & 2.18& 1.51& 3.42& 1.91&                    6.75& 6.10& 10.39& 7.40\\
				R-CE       & 2.17& 1.49& 3.42& 1.89&                    6.42& 5.76& 10.05& 7.09\\
				DeCA       & 2.20& 1.48& 3.54& 1.91&                    6.67& 6.12& 10.55& 7.52\\
				DCF        & 2.19& 1.51& 3.41& 1.91&                    6.80& 6.15& 10.53& 7.48\\
				PLD        & 1.97& 1.27& 3.20& 1.69&                    7.23& 6.30& 11.12& 7.97\\
				\cmidrule(r){1-9} 
				FO         & 2.24& 1.54& 3.32& 1.89&                    7.22& 6.65& 11.12& 8.04\\
				TPSC       & 2.45& 1.67& 3.80& 2.09&                    8.28& 9.01& 11.70& 9.93\\
				TPSC-FO    & \textbf{2.59}& \textbf{1.80}& \textbf{3.87}& \textbf{2.20}&    \textbf{8.91}& \textbf{9.75}& \textbf{12.59}& \textbf{10.74}\\
				\bottomrule
			\end{tabular}
	}}

\end{table}

\subsection{Performance Comparison (RQ1)}

We present the performance of TPSC-FO compared to other baselines on real-world and noisy datasets in Tables II and III, with the following observations:

\begin{itemize}

\item TPSC-FO achieves the best results across all metrics on all datasets, with both TPSC and FO modules enhancing the performance of LightGCN (RNS). We attribute this to: (1) the TPSC module, which effectively converts false negatives into positive samples, and (2) the FO module, which mitigates noise in positive sample embeddings, enabling more accurate learning of user preferences.

\item Denoising methods typically underperform negative sampling methods, as overall loss-based denoising struggles to distinguish noisy interactions from normal ones, weakening positive supervision signals and reducing model performance, consistent with \cite{29} experimental results. In contrast, negative sampling methods enhance model discrimination with hard negative samples while maintaining positive supervision signals, leading to performance improvements. TPSC-FO strengthens positive supervision signals via TPSC and denoises through FO without weakening these signals, achieving greater performance gains.

\item The performance improvement achieved by TPSC-FO becomes more pronounced as the dataset scale and density increase. This is because larger and denser datasets provide richer topological structures in the user–item bipartite network, enabling differential community detection algorithm to identify communities more accurately and consistently. This leads to a higher recall of potential false negatives in the candidate set $Q_u$. Although the enlarged candidate set inevitably includes more noisy samples, the subsequent personalized noise filtration based on user-specific similarity thresholds effectively removes noisy items while retaining truly false negatives. In contrast, smaller and sparser datasets exhibit limited topological connectivity, which restricts community detection and leads to relatively fewer identified false negatives. Consequently, the performance improvement brought by TPSC-FO, though still significant, is relatively smaller on small-scale datasets.

\item Compared to real-world datasets, TPSC-FO yields greater performance improvements on synthetic datasets. For example, on the Amazon-beauty,  Epinions, and corresponding synthetic datasets, Recall@20 improvements over the strongest baseline increase from 5.45\% to 6.61\% and from 8.97\% to 11.42\%, respectively. This further demonstrates that TPSC-FO effectively identifies false negatives.

\end{itemize}

\begin{figure}[t]
	\centering

	\includegraphics[width=\linewidth]{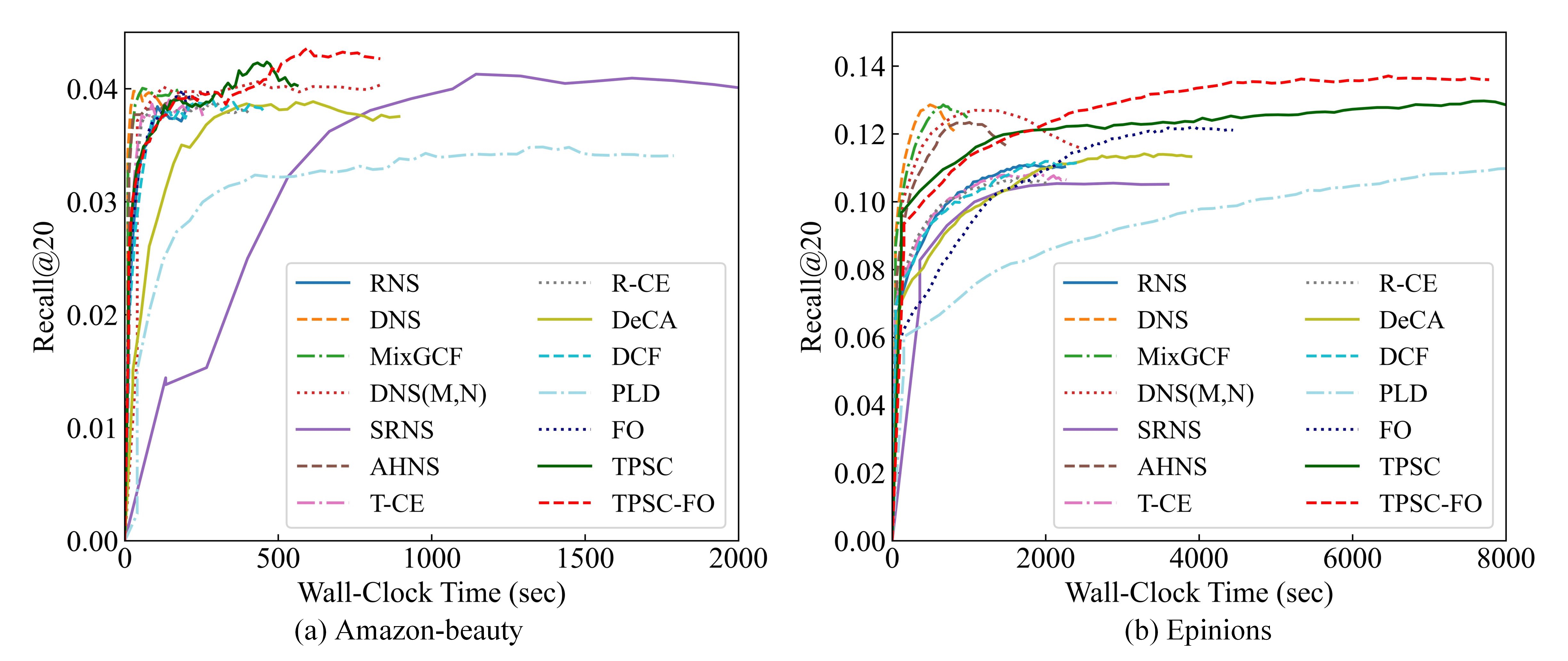}
	\caption{Recall@20 vs. wall-clock time (in seconds).}
	
\end{figure}

\begin{figure*}[t]
	\centering

	\includegraphics[width=\linewidth]{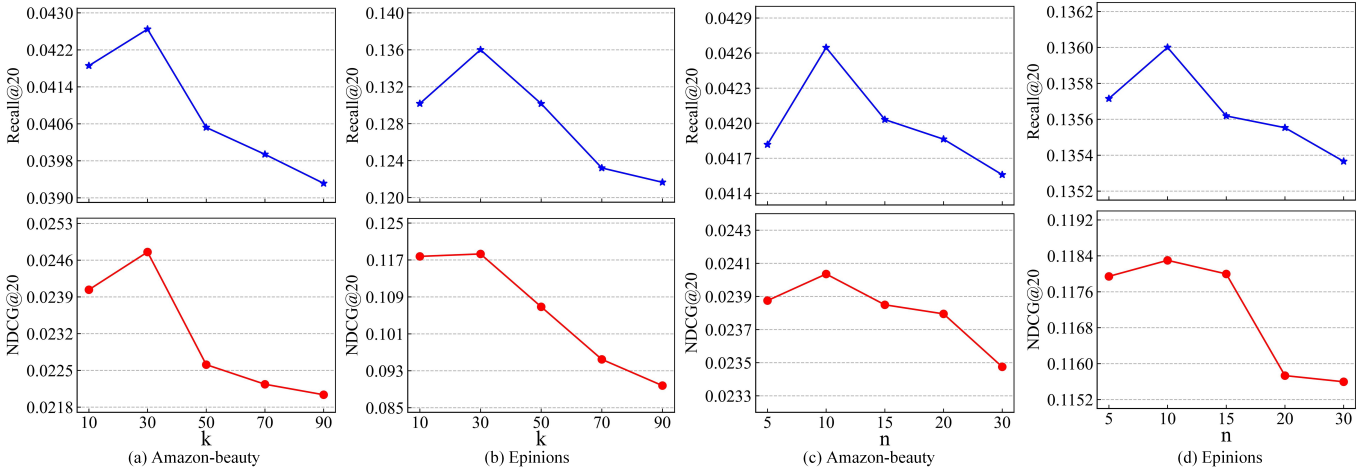}
	
	\vspace{-0.2cm}
	
	\caption{The impact of $k$ and $n$.}
	
		\vspace{-0.2cm}
	
\end{figure*}

Besides effectiveness, we analyzed the efficiency of TPSC-FO. We first measured TPSC’s runtime on the Amazon-beauty, Amazon-home, Epinions, Gowalla, and Tmall datasets, which were 30s, 71s, 106s, 297s, and 482s. Notably, TPSC requires only a single construction before model training, demonstrating high feasibility. We then compared the average per-epoch training time on the Amazon-beauty, Amazon-home, Epinions, Gowalla, and Tmall datasets after incorporating FO, which increased from 1.09s, 2.95s, 11.29s, 28.61s, 208.09s to 1.12s, 3.02s, 11.90s, 29.09s, 210.73s. This indicates that FO’s computational overhead constitutes only a minimal fraction of the training time. Additionally, we compared TPSC-FO’s training efficiency with baselines in Fig.~5, revealing that TPSC-FO converges to higher performance in a short period. The aforementioned experimental results validate the superior efficiency and effectiveness of TPSC-FO.

\begin{figure}[t]
	\centering
	
		\vspace{-0.2cm}
	
	\includegraphics[width=\linewidth]{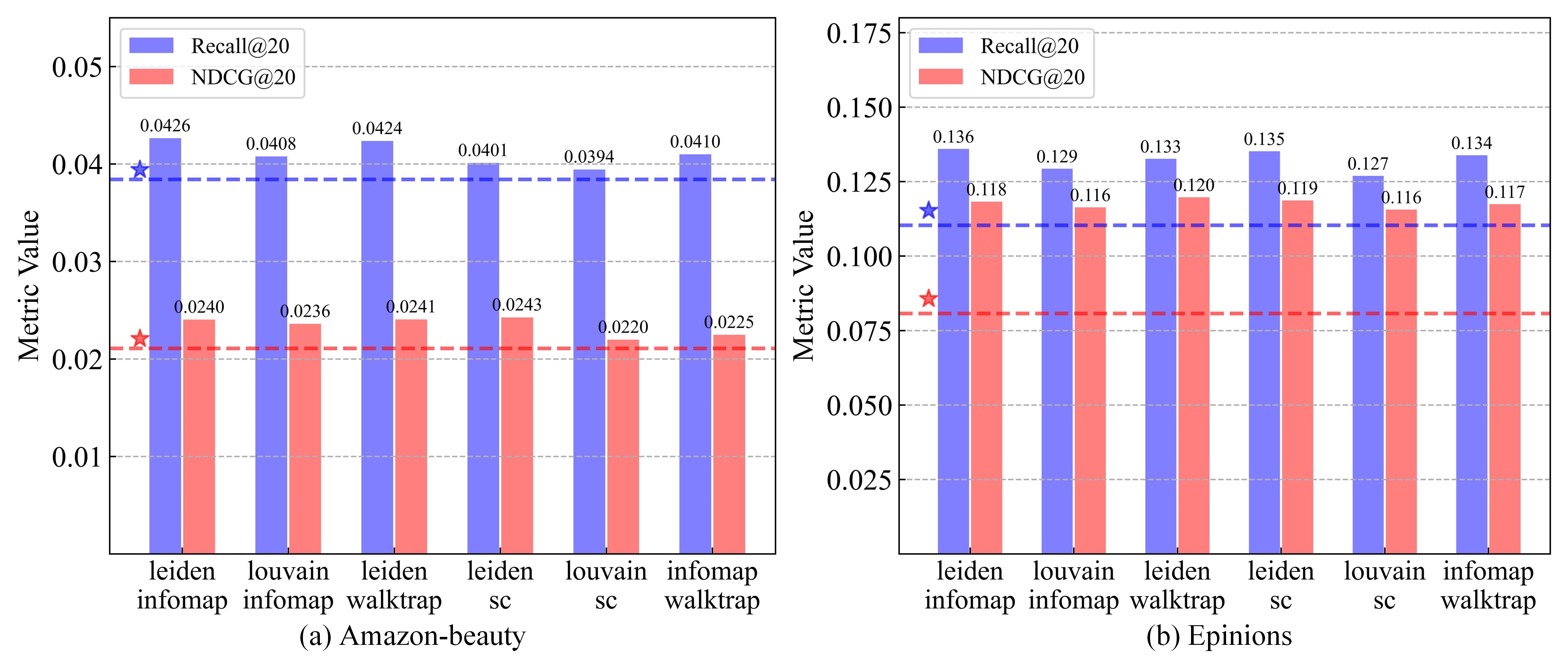}
	\caption{The impact of community detection combinations.}
	
\end{figure}

\subsection{Parameter Analysis (RQ2)}

In this section, we show the impact of the quantile $ k $, which determines the effectiveness threshold for false negatives, and the size of the neighborhood positive sample set $ n $ on model performance. The parameter $ k $ takes values from \{10, 30, 50, 70, 90\}, and $ n $ from \{5, 10, 15, 20, 30\}. When analyzing one parameter, the other is held constant.

Fig.~6(a)-6(b) illustrates the impact of $k$ on model performance. It can be observed that as $k$ increases, model performance generally shows a trend of initially rising and then declining. This occurs because a smaller $k$ includes noisy samples in the positive sample set, which degrades performance, whereas a larger $k$ excludes some potential positive samples, leading to a higher risk of false negatives and thus also affecting performance. Experimental results show that when $k$ = 30, TPSC-FO achieves superior performance across various datasets.

\begin{figure}[t]
	\centering
	
		\vspace{-0.2cm}
	
	\includegraphics[width=\linewidth]{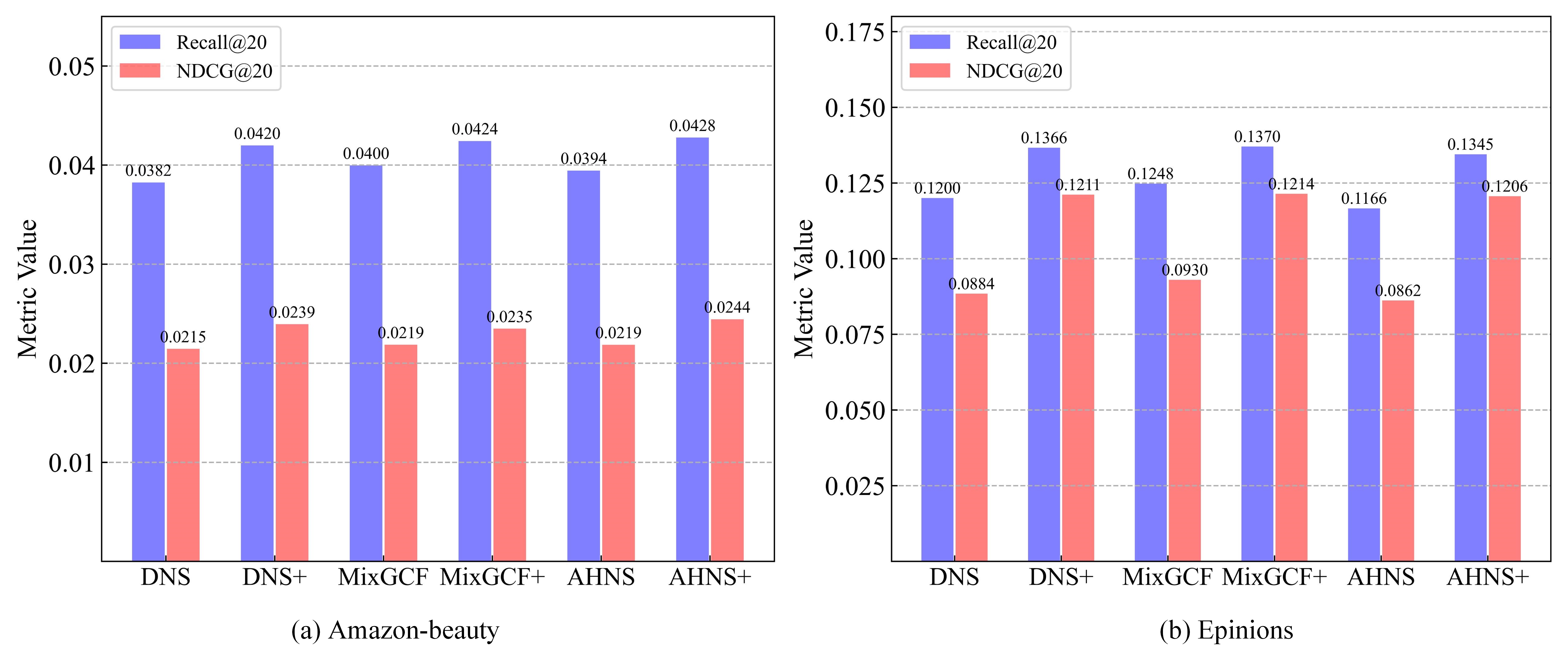}
	\caption{Performance comparison of integrating TFNS with different negative sampling methods.}
\end{figure}

Fig.~6(c)-6(d) show $n$'s impact on model performance. As $n$ increases, performance exhibits an overall trend of initially improving and then declining. This is because, when $n$ is small, noisy samples significantly disrupt the extraction of common features from the neighborhood sample set, thereby impairing feature optimization. Conversely, when $n$ is large, although common features become more robust, over-smoothing occurs, degrading performance. Experimental results show that TPSC-FO achieves good performance across different datasets when $n$ = 10.

\subsection{Robustness Analysis (RQ3)}

In this section, we examine how different community detection algorithms and node embedding methods adopted in the TPSC module affect the overall performance of TPSC-FO.

We selected Leiden, Infomap, Louvain, Walktrap, and Spectral Clustering for combination, examining the effects of various differential community strategies on model performance. As shown in Fig.~7, the horizontal line represents the performance of RNS (LightGCN). It can be observed that on the Amazon-Beauty and Epinions datasets, the fluctuations of Recall@20 and NDCG@20 are within (0.0017, 0.0013) and (0.0054, 0.0021), respectively, and that all combinations based on different community detection algorithms consistently improve model performance, demonstrating the robustness of TPSC-FO. This robustness primarily stems from the subsequent personalized noise filtration, which filters out noisy samples, and the neighborhood-guided feature optimization, which suppresses noisy information in feature representations. Combined with Fig.~2, we observe that community detection algorithms with higher false negative identification ratios yield greater performance improvements.

Considering that node embedding learning methods are only part of TPSC module and require balancing time complexity, we selected MF-based NMF and SVD for comparison. As shown in Table IV, all three methods effectively learn node embeddings in a short time, with SVD being more efficient on smaller datasets and ALS more efficient on larger and sparser datasets. Moreover, ALS outperforms NMF and SVD in performance, benefiting from its iterative optimization tailored for sparse data (e.g., weighted regularization), which enables more accurate node embedding learning.

\begin{table}[t]
	\centering
	
	\setlength{\abovecaptionskip}{0.2cm}
	\setlength{\belowcaptionskip}{-0.2cm}
	
	\caption{Impact of node embedding methods in TPSC.}
	\resizebox{\linewidth}{11mm} {
		\setlength{\tabcolsep}{1.2mm}{
			\begin{tabular}{c|ccc|ccc}
				\toprule
				\multirow{1}{*}{Dataset} & \multicolumn{3}{c|}{Amazon-beauty} & \multicolumn{3}{c}{Amazon-home} \\ 
				\cmidrule(r){1-1}\cmidrule(r){2-4} \cmidrule(r){5-7}
				\multirow{1}{*}{Method}& R@20 & N@20 & Running time & R@20 & N@20 & Running time  \\
				\midrule
				ALS        & \textbf{4.26}& \textbf{2.40}& 0.61s& \textbf{2.09}&       \textbf{1.10}& \textbf{0.81s}\\
				SVD        & 4.04& 2.34& \textbf{0.57s}& 2.06&       1.06& 1.87s\\
				NMF        & 3.94& 2.33& 4.46s& 1.98&       1.04& 14.15s\\
				\bottomrule
			\end{tabular}
	}}
\end{table}

\begin{table}[t]
	\centering

	\setlength{\abovecaptionskip}{0.2cm}
	\setlength{\belowcaptionskip}{-0.2cm}
	
	\caption{Performance comparison integrated with MF (\%).}
	\resizebox{\linewidth}{24mm} {
		\setlength{\tabcolsep}{1.2mm}{
			\begin{tabular}{c|cccc|cccc}
				\toprule
				\multirow{1}{*}{Method} & \multicolumn{4}{c|}{Amazon-beauty} & \multicolumn{4}{c}{Epinions} \\ 
				\cmidrule(r){2-5} \cmidrule(r){6-9}
				& R@10 & N@10 & R@20 & N@20 & R@10 & N@10 & R@20 & N@20 \\
				\midrule
				RNS        & 1.32& 0.94& 2.16& 1.20& 5.68& 5.20& 9.04& 6.41\\
				DNS        & 1.64& 1.17& 2.70& 1.50& 5.83& 5.49& 8.92& 6.58\\
				DNS(M,N)   & 1.72& 1.23& 2.81& 1.56& 5.62& 5.43& 8.64& 6.46\\
				SRNS       & 1.70& 1.22& 2.76& 1.55& 5.76& 5.44& 8.88& 6.54\\
				AHNS       & 1.67& 1.17& 2.57& 1.46& 6.20& 5.86& 9.58& 7.04\\
				\cmidrule(r){1-9} 
				T-CE       & 1.01& 0.74& 1.48& 0.89& 5.91& 5.51& 8.94& 6.57\\
				R-CE       & 1.08& 0.79& 1.65& 0.97& 5.28& 4.74& 8.46& 5.89\\
				DeCA       & 1.26& 0.89& 2.11& 1.16& 5.50& 5.16& 8.79& 6.32\\
				DCF        & 1.06& 0.77& 1.60& 0.94& 5.37& 4.85& 8.80& 6.09\\
				PLD        & 1.36& 0.90& 2.19& 1.17& 5.41& 4.72& 8.44& 6.03\\
				\cmidrule(r){1-9} 
				TPSC-FO    & \textbf{1.98}& \textbf{1.54}& \textbf{3.03}& \textbf{1.86}& \textbf{8.79}& \textbf{9.80}& \textbf{12.27}& \textbf{10.67}\\
				\bottomrule
			\end{tabular}
	}}
	
	\vspace{-0.2cm}
	
\end{table}

\subsection{Integrated with Other CF Models and Negative Sampling Methods (RQ4)}

Table V presents the experimental results of TPSC-FO compared to baselines in MF-based CF model \cite{53}. We omitted MixGCF, as it is designed based on graph neural network aggregation processes. It can be observed that TPSC-FO significantly outperforms other baselines, owing to our approach of starting from the essence of false negatives to construct a high-quality positive sample set. This explicit guidance signal enables models with limited representation capabilities to accurately learn user preferences. We integrate TPSC-FO with three popular negative sampling methods (DNS, MixGCF, and AHNS) yielding DNS+, MixGCF+, and AHNS+. Specifically, we use the TPSC module to construct a topology-aware positive sample set and the FO module to denoise positive sample embeddings, collaboratively training the recommendation model with negative sample embeddings from these negative sampling methods. As shown in Fig.~8, TPSC-FO significantly enhances the performance of these methods, indicating its ability to effectively mitigate the risk of false negatives and improve the reliability of negative samples obtained. The aforementioned experimental findings collectively confirm the strong applicability of TPSC-FO.

\begin{table}[t]
	\centering
	\caption{Case study of user~3930 showing community assignments and ranking comparison.}
	
	\begin{tabular}{c|cc|ccc}
		\toprule
		\multirow{2}{*}{Node ID} & 
		\multicolumn{2}{c|}{Community Label} & 
		\multicolumn{3}{c}{Rank in Recommendation List} \\
		\cmidrule(lr){2-3} \cmidrule(lr){4-6}
		& Leiden & Infomap & Original & TPSC & TPSC-FO \\
		\midrule
		3930 & 0 & 1 & -- & -- & -- \\
		\midrule
		10702 & 0 & 1 & 128 & 5 & \textbf{4} \\
		9853  & 0 & 1 & 45 & 4 & \textbf{2} \\
		10154 & 0 & 1 & 233 & 6 & \textbf{3} \\
		9036  & 0 & 1 & 281 & 37 & \textbf{29} \\
		\bottomrule
	\end{tabular}
	\label{tab:case-study}
\end{table}

\begin{figure}[!]
	\centering
	\includegraphics[width=\linewidth]{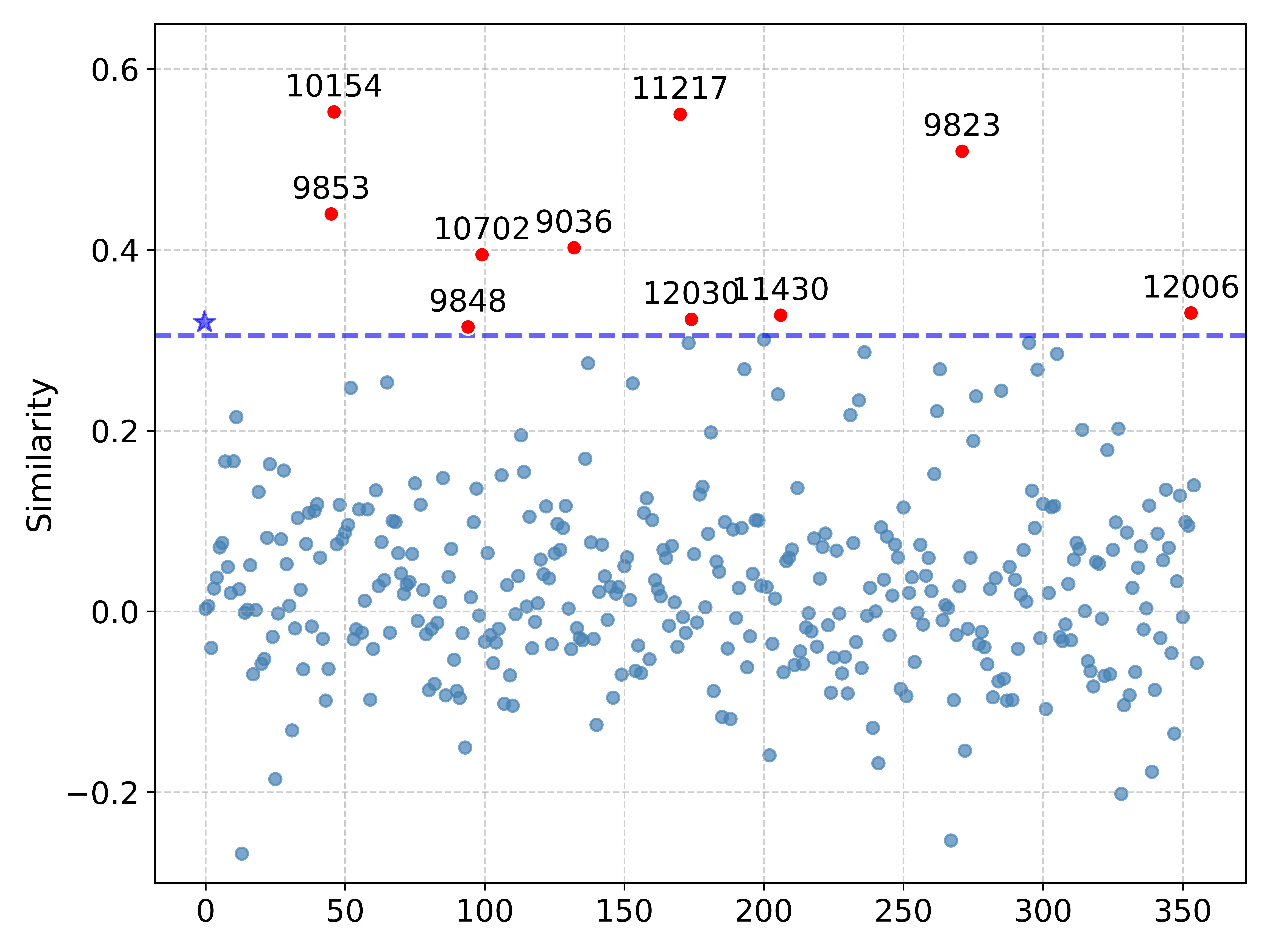}
	\caption{Personalized noise filtration results for user~3930.}
		\label{fig:case-threshold}
\end{figure}

\subsection{Case Study}

To further highlight the practical advantages of the TPSC-FO, we present a user-level case study. In the Amazon-Beauty dataset, we select the user node with ID 3930 and randomly remove 20\% of their historical interactions (items 10702, 9853, 10154, and 9036), which are treated as false negatives. We first perform differential community detection strategy to the user–item interaction network constructed from this dataset. As shown in the left part of Table~\ref{tab:case-study}, both Leiden and Infomap assign user~3930 and the four flipped items to the same community, confirming that the differential community detection strategy can effectively identify false negatives. In the consensus of the two partitions, 372 items are assigned to the same community as user~3930. We then apply personalized noise filtration to these items. As illustrated in Fig.~\ref{fig:case-threshold}, the blue dashed line represents the user-specific threshold computed for user~3930. The results show that this mechanism effectively filters out noisy samples introduced by the forced merging of small-scale interest communities in community detection, while successfully identifying all four flipped items as valid false negatives and identifying six additional false negatives in the negative sample space. Finally, we train LightGCN using the original data, TPSC, and TPSC-FO, and compare the rankings (lower is better) of the four flipped items in the recommendation list of user~3930. To prevent data leakage, the four flipped items identified by TPSC-FO are removed from the training set, while the remaining six identified potential false negatives are retained, none of which appear in the validation or test sets.
As shown in Table~\ref{tab:case-study}, the right part presents the ranks of the four flipped items under the Original, TPSC, and TPSC-FO settings.
It can be observed that TPSC effectively improves the rankings of these false negatives in the Top-$K$ recommendation list. 
This indicates that converting the six identified false negatives into positive samples for user~3930 enables the model to better capture the user’s true preferences, thereby substantially improving model performance. 
Furthermore, by introducing the FO module, TPSC-FO effectively mitigates noise within positive sample embeddings and achieves further improvements in recommendation accuracy.

\section{Conclusion}
In this paper, we study negative sampling in implicit CF from a new perspective that false negatives can serve as positive samples to guide model training. We innovatively employ topological community features to identify false negatives, constructing a topology-aware positive sample set that effectively transforms interference signals into positive supervision signals. Additionally, we design a neighborhood-guided feature optimization module to mitigate noise in positive sample embeddings. Extensive experiments demonstrate that TPSC-FO significantly outperforms other baseline models in implicit CF and can integrate with other negative sampling methods to further enhance their performance. This introduces a novel research direction for negative sampling strategies, focusing on precisely identifying false negatives to construct high-quality positive sample sets, thereby further improving implicit CF model performance.

\bibliographystyle{IEEEtran}
\bibliography{citiation}

\vfill

\end{document}